\documentclass[12pt]{report}
\usepackage[dvips]{graphicx}
\newcommand{\sptwo}{1.4}

\newcommand{\doublespace}{\edef\baselinestretch{\sptwo}\Large\normalsize}

\begin{document}
\doublespace
\begin{center}
{\bf Collective Excitations  of  Strongly Interacting Fermi Gases of Atoms in a Harmonic Trap}\\
\renewcommand\thefootnote{\fnsymbol{footnote}}
{Yeong E. Kim \footnote{ e-mail:yekim$@$physics.purdue.edu} and
Alexander L. Zubarev\footnote{ e-mail: zubareva$@$physics.purdue.edu}}\\
Purdue Nuclear and Many-Body Theory Group (PNMBTG)\\
Department of Physics, Purdue University\\
West Lafayette, Indiana  47907\\
\end{center}

\begin{quote}
The zero-temperature properties of a dilute two-component Fermi gas in the
BCS-BEC crossover are investigated. On the basis of a generalization of the
 Hylleraas-Undheim method, we construct rigorous upper bounds to the collective
 frequencies for the radial and the axial breathing mode of the Fermi gas
 under harmonic confinement in the framework of the
 hydrodynamic theory. The bounds are compared to experimental data for trapped
 vapors of $^6Li$ atoms.
\end{quote}

\vspace{5mm}
\pagebreak

The recently reported ultracold trapped Fermi gases with tunable atomic
 scattering length [1-19] in the vicinity of a Feshbach resonance stimulated a
 large number of theoretical investigations. In this letter the dynamics of
 strongly interacting dilute Fermi gases (dilute in the
sense that
 the range of interatomic potential is small
compared with inter-particle spacing
) is investigated in the framework of the hydrodynamic theory [20-29].

Instead of using  the scaling approximation of Refs[20-22,25,27], the polytropic 
approximation of Refs.[24,28] or the perturbative approximation of Ref.[26], 
we will construct a rigorous upper bounds to the collective frequencies.

Our starting point is  the variational
formulation of
the Kohn-Sham time-dependent theory for a dilute two-component Fermi gas in a 
trap potential $V_{ext}(\vec{r})=(m/2)(\omega_{\perp}^2 (x^2+y^2)+
\omega_z^2 z^2)$
$$
\delta \int dt<\Psi|i\hbar \partial_t-H|\Psi> = 0,
\eqno{(1)}
$$
where $|\Psi>$ is a product of two Slater determinants, one for each internal
 state built up by the Kohn-Sham orbitals $\psi_i$,  and $H=T+U$ is 
Hamiltonian in the local density approximation (LDA).

 We consider two approximations:

(i) local transform $\psi_i\approx \phi_i \exp (i\chi)$,
where $\psi_i$ and $\chi$ are real functions,

\noindent
and

(ii) $<\phi|T|\phi>\approx (1/N) \int (t_{TF}(n)+t_W(n))n(\vec{r},t)d^3 r,$

where $n$ is the density, normalized to the total number of atoms $N$,
$\int n d^3 r=N$, $|\phi>$ is the product of two Slater determinants 
built on $\phi_i$
alone,
$t_{TF}(n)=(3 \pi^2)^{2/3} (3 \hbar^2/(10 m)) n(\vec{r},t)^{2/3}$ is the
Thomas-Fermi (TF)  kinetic energy density, and
$t_W(n)=(\hbar^2/(8m))(\nabla n)^2/n$ is the  von Weizs\"{a}cker kinetic energy
density.

Using the approximations (i) and (ii),
we obtain from Eq.(1) [20-22]
$$
i\hbar \frac{\partial \psi}{\partial t}=-\frac{ \hbar^2}{2 m} \nabla^2 \psi
+V_{ext} \psi+V_{xc}\psi,
\eqno{(2)}
$$
where $\psi=\sqrt{n} \exp (i \chi)$, 
$V_{xc}(\vec{r},t)=[\frac{d n \epsilon(
n)}{d n}]_{n=n(\vec{r},t)}$ and $\epsilon(n)$ is the ground state energy per particle of a uniform system.
The only difference from equations holding for bosons [30,31] is given by density 
dependence of $\epsilon(n)$.

It is useful to rewrite Eq.(2) in a  form
$$
\frac{\partial n}{\partial t}+\nabla (n \vec{\mbox{v}})=0,
\eqno{(3)}
$$
$$
\frac{\partial \vec{\mbox{v}}}{\partial t}+\frac{1}{m}\nabla (V_{ext}+
\frac{d(n \epsilon(
n))}{d n}+\frac{1}{2} m \mbox{v}^2-\frac{\hbar^2}{2 m} \frac{1}{\sqrt{n}}
 \nabla^2 \sqrt{n})=0,\eqno{(4)}
$$
where $\vec{\mbox{v}}=(\hbar/m) \nabla \chi$ is the velocity field.

 It was shown in Refs.[20,21] that for experimental conditions of
 Refs.[14,16,17]
the quantum pressure term in Eq.(4)  can be neglected. For the reminder of this Letter we
 will use this hydrodynamic approximation.

For the negative S-wave scattering length between the two fermionic species,
 $a<0$,
in the low-density regime, $k_F\mid a \mid \ll 1$, the ground state energy per
particle , $\epsilon(n)$, is well represented by an expansion in power of
$k_F \mid a \mid$ [32]
$$
\epsilon(n)=2 E_F[\frac{3}{10}-
\frac{1}{3 \pi} k_F \mid a \mid+0.055661 (k_F\mid a \mid)^2
-0.00914 (k_F\mid a \mid)^3+...],
\eqno{(5)}
$$
where $E_F=\hbar^2 k_F^2/(2 m)$ and $k_F=(3 \pi^2 n)^{1/3}$.
In the opposite regime, $a\rightarrow - \infty$
(the Bertsch many-body problem, quoted in
Refs.[33]), $\epsilon(n)$ is proportional to that of the non-interacting Fermi
 gas
$$
\epsilon(n)=(1+\beta)\frac{3}{10} \frac{\hbar^2 k_F^2}{m},
\eqno{(6)}
$$
where  the universal parameter  $\beta$ [10]  is estimated to be $\beta=-0.56$
 [34]. The universal limit [10,34-38] is valid at least in the case where the
width of the Feshbach resonance is large compared to the Fermi energy as in
 the cases of $^6Li$ and $^{40}K$.

 In the $a\rightarrow +0$ limit the system reduces to the
dilute Bose gas of dimers [39]
$$\epsilon(n)=E_F(-1/(k_F a)^2+a_m k_F/(6 \pi)+...,
)
\eqno{(7)}
$$
where $a_m$ is the boson-boson scattering length, $a_m \approx 0.6 a$ [40].

A simple interpolation of the form $\epsilon(n)\approx
E_F P(
k_F  a)$ with a smooth function $P(x)$ was considered in several papers.
In Ref.[20] it has  been proposed a [2/2] Pad\'{e} approximant for the function
$P(x)$ for the  the case of negative $a$
$$
P(x)=\frac{3}{5}-2\frac{\delta_1\mid x \mid+\delta_2 x^2
}{1+\delta_3\mid x\mid+\delta_4 x^2},
\eqno{(8)}
$$
where $\delta_1=0.106103$, $\delta_2=0.187515$, $\delta_3=2.29188$,
$\delta_4=1.11616$.
Eq.(8) is constructed to reproduce the first four terms of the expansion (5) in
 the low-density regime and
  also  to exactly reproduce
 results of
 the recent Monte Carlo calculations [34], $\beta=-0.56$, in the  unitary limit,
$k_F a \rightarrow -\infty$.

 For the positive $a$ case ( the interaction is strong enough to form bound
 molecules
 with energy $E_{mol}$)  we have considered in Ref.[21]  a [2/2] Pad\'{e}
 approximant
$$
P(x)=\frac{E_{mol}}{2 E_F}+\frac{\alpha_1 x+\alpha_2 x^2}{1+\alpha_3 x+\alpha_4
x^2},
\eqno{(9)}
$$
where
 parameters $\alpha$ are fixed by two continuity conditions at large $x$,
$1/x\rightarrow 0$, and by two continuity conditions at small $x$,
 $\alpha_1=0.0316621$, $\alpha_2=0.0111816$, $\alpha_3=0.200149$,
and $\alpha_4=0.0423545$.

In Ref.[41] a Pad\'{e} approximation  has been considered   for the
 chemical potential. Authors of Ref.[28] have used a model for $P(x)$,
 interpolating the Monte Carlo results of Ref.[36] across the unitary limit
 and limiting behaviors for small $|x|$. We note here also the BCS mean-field 
calculations of Ref.[27]

The hydrodynamic equations after linearization take the form
$$
\frac{\partial^2}{\partial t^2} \delta n+\frac{1}{m} \nabla (
n \nabla (\frac{d^2(n \epsilon(n))}{d n^2} \delta n))=0,
\eqno{(10)}
$$
where $\delta n(\vec{r},t)$ is the change in the density profile with respect 
to the equilibrium configuration. If we consider oscillations with time
 dependence $\delta n \propto \exp(i \omega t)$, Eq.(10) can be reduced to
a  Hermitian equation [26] 
$$
-\omega^2 [\frac{d^2(n \epsilon(n))}{d n^2}]^{-1} f=
\frac{1}{m} \nabla(n\nabla f),
\eqno{(11)}
$$
where $f=\frac{d^2(n \epsilon(n))}{d n^2}\delta n$, and the
 equilibrium density, $n$, is given by equation
$$
\mu=V_{ext}+\frac{d(n \epsilon(n))}{d n},
\eqno{(12)}
$$
where $\mu$ is the chemical potential, in the region where $n(\vec{r})$ is
 positive and $n(\vec{r})=0$ outside this region.

The Hilleraas-Undheim method [42] for upper bounds to eigenenergies of excited 
states of atoms can be generalized to the case of Eq.(11). 
Defining the functional $I[\chi]$ by

 $$
I[\chi]=-(1/m) \int \chi \nabla (n \nabla \chi) d^3 r/\int\chi 
[\frac{d^2(n \epsilon(n))}{d n^2}]^{-
1}\chi d^3 r,
$$
 we get $I[f]=\omega^2,$ where $f$ is the solution of Eq.(11).

Let the sequence of
 eigenvalues of Eq.(11) be denoted by $\omega_1^2 \leq \omega_2^2 \leq ...$
and let $\chi_1, \chi_2, ..., \chi_q$ form a set of $q$ linear independent
 functions. 
Introducing the function $\chi=\sum_{r=1}^q c_r \chi_r$, where $c_1,c_2, ... c_q$ are $q$ variable parameters, we see that the functional $I$ is stationary if $\partial I/\partial c_r=0$.
 Then the values of the $q$ roots, $\tilde{\omega}_1^2\leq
 \tilde{\omega}_2^2 \leq \tilde{\omega}_3^2 \leq ...\leq \tilde{\omega}_q^2$,
of the secular equations
$$
det[-\frac{1}{m} \int \chi_p \nabla (n \nabla \chi_s) d^3 r
-\tilde{\omega}^2 \int\chi_p [\frac{d^2(n \epsilon(n))}{d n^2}]^{-1}\chi_s d^3 r]=0
\eqno{(13)}
$$
provide upper bounds to the $q$ lowest eigenvalues of Eq.(11)
$
\tilde{\omega}_r^2\geq \omega_r^2 $ for $r=1,2, ..., q$.
It follows that as extra terms are added to the expansion of $\chi$, 
a given root of the secular equation $\tilde{\omega}_r$ is decreased.

For the most interested case of $M=0$ modes, we can put in Eq.(13) $q=3$,$~$ $\chi_1=1$,
$~$ $\chi_2=(x^2+y^2)$ and $\chi_3=z^2$, which give
$$
\frac{\omega_{\pm}^2}{\omega_{\perp}^2}=\frac{\eta \pm \sqrt{
\eta^2-8 \lambda^2 \zeta (5 \zeta-9)}}
{(5 \zeta-9)},
\eqno{(14)}
$$
where $\eta= (3+4 \lambda^2) \zeta - (3+6 \lambda^2)$,
 $\zeta=I_0 I_4/I_2^2,$
$I_l=\int \tilde{x}^l n(\tilde{x}) d\tilde{x}$, 
$\tilde{x}=\sqrt{x^2+y^2+\lambda^2 z^2}$, $\lambda=\omega_z/\omega_{\perp},$ and $\pm$ signs refer to the
 transverse and axial mode, respectively.

To calculate $\zeta$, we have used the following expansion
$$
n(\vec{r})\approx (1-\beta V_{ext}(\vec{r}))^{1/(2-p)}
\sum_{i=0}^{l-1} c_{i} [V_{ext}(\vec{r})]^i,
\eqno{(15)}
$$
where parameters $\beta$, $p$ and $c_i$ are fixed by requiring that
 $n(\vec{r})$ must satisfy a variational principle
 $\delta \int n (V_{ext}+\epsilon(n))d^3 r=0$ with a subsidiary condition
 $\int n d^3 r=N$.

From the Table 1 one can see a very fast convergence for $\zeta$ and consequently for $\omega_{\pm}$, the first term of the expansion (15) has accuracy $10^{-3}$ [43], while the sum of the first three terms has accuracy $10^{-4}$.

It is easy to show  that our  upper bounds, Eq. (14), give exact
 solutions for frequencies of the breathing modes for the polytropic equation
 of state, $\epsilon(n)\approx n^{\gamma}$. 

The hydrodynamic equation is expected to be applicable
for describing the macroscopic excitations of the system up to energies of the
 order of the energy gap, $\Delta$, needed to break-up a Cooper pair. But for
 the trapped gas, $\Delta$ is a function of $\vec{r}$ ($\Delta$ decreasing when
 we
 go away from the center). It is natural to assume that
the  condition of the
applicability of hydrodynamics to describe the macroscopic excitations of the
system at $T=0$ is [22,44]      
$$
\frac{\hbar \Omega}{\tilde{\Delta}}\ll 1,
\eqno{(16)}
$$
where $\Omega$ is the frequency of the macroscopic excitations
and the mean energy gap is given by  
$\tilde{\Delta}=\int n(\vec{r}) \Delta(\vec{r}) d\vec{r}
/N$.
To calculate $\tilde{\Delta}$ we have used results of Refs.[45,46].
The predictions of Eq.(14) for the radial breathing mode frequency of the
 cloud of $^6Li$ atoms are shown in Fig.1. For $\epsilon(n)$ we have used
 two approximations, the Pad\'{e} [2/2] approximation of Refs.[20,21] and the
 parameterization of Ref.[28]. It can be
 seen a very small difference between these two approximations.

 In Fig.1, we have also compared the hydrodynamic predictions with experimental 
data [17]. We have used Ref.[47] to convert $a$  to the magnetic field $B$.
 Near the unitary limit  there
 is a very good agreement with experimental data [17].  
  We note here that two
 experimental results [14,17] and [16] for $\omega_+$ are still 
about  10\%  in disagreement  with
 each other, which is not fully understood yet.
 At  $[N^{1/6} a/a_{ho}]^{-1}=-1.26$ the hydrodynamic upper bound begins to be below
 the  sharp increased measured frequency that confirms a breakdown of
 hydrodynamic  theory[17].

In Fig.2, the calculated hydrodynamic upper bounds to the axial breathing mode 
frequency of the cloud of $^6Li$ is compared with experimental data [16] in the
 BCS-BEC crossover region. It can be seen that the difference between the  two
 approximations for  $\epsilon(n)$ is practically negligible and both
 approximations give a very good agreement between calculations and experimental data [16] for $[N^{1/6} a/a_{ho}]^{-1}\geq -1.14$. However at 
$[N^{1/6} a/a_{ho}]^{-1}=-1.31$ the hydrodynamic upper bound begins to be below the
 measured frequency that  indicates, as in the radial  mode case,
 breakdown of hydrodynamics, even though  for the
 axial mode case the ratio between the collective energy and the gap energy, Eq.(16)  is very small,
 $\hbar \omega_-/\tilde{\Delta}\approx 0.01$, at $[N^{1/6} a/a_{ho}]^{-1}=-1.31$ [48]. This breakdown of 
hydrodynamic theory  may
be related to the finite-temperature effects.

In summary, we have  constructed the rigorous upper bounds to the collective
 frequencies of the Fermi gas under harmonic confinement in the framework of
 the hydrodynamic
 theory. The bounds are compared to experimental data on confined vapors of
 $^6Li$ atoms.
It is shown that the 
 hydrodynamic upper bound begins to be below the measured frequency at 
$[N^{1/6} a/a_{ho}]^{-1}\approx -1.3$, although the ratio between the collective energy and the gap energy, Eq.(16), for the axial mode is very small. 
 
We thank J. Thomas for stimulating this work and for sharing with us the
 experimental results prior to publication. We also thank R. Grim and
 M. Bartenstein for providing us with the updated experimental data.
\pagebreak

Table 1. $\zeta$ in the BCS region as a function of the dimensional parameter
 $X=(N^{1/6} a/a_{ho})^{-1}$ and the number of terms, $l$, 
in the expansion (15).
  The [2/2] Pad\'{e}                                                     
approximation
 of Refs.[20,21] is used for the energy per particle $\epsilon(n)$.\\

\begin{tabular}{llllll}
\hline\hline
X
&$l=1$
&$l=2$
&$l=3$
&$l=4$
&$l=5$ \\ \hline
-0.1
&2.2587
&2.2572
&2.2577
&2.2575
&2.2576 \\ \hline
-0.2
&2.2658
&2.2637
&2.2642
&2.2640
&2.2641 \\ \hline
-0.4
&2.2753
&2.2732
&2.2735
&2.2735
&2.2735 \\ \hline
-0.6
&2.2801
&2.2789
&2.2787  
&2.2789
&2.2788 \\ \hline
-0.8
&2.2821
&2.2818
&2.2814
&2.2816
&2.2815 \\ \hline
-1.1
&2.2822
&2.2830
&2.2825
&2.2827
&2.2826 \\ \hline
-1.4
&2.2810
&2.2825
&2.2819
&2.2821
&2.2820 \\ \hline
-1.7
&2.2792
&2.2811
&2.2805
&2.2807
&2.2806 \\ \hline
-2.0
&2.2774
&2.2795
&2.2789
&2.2791
&2.2790 \\ \hline
-3.0
&2.2720
&2.2742
&2.2738
&2.2739
&2.2739 \\ \hline \hline
\end{tabular}

\pagebreak

\begin{figure}[ht]
\includegraphics{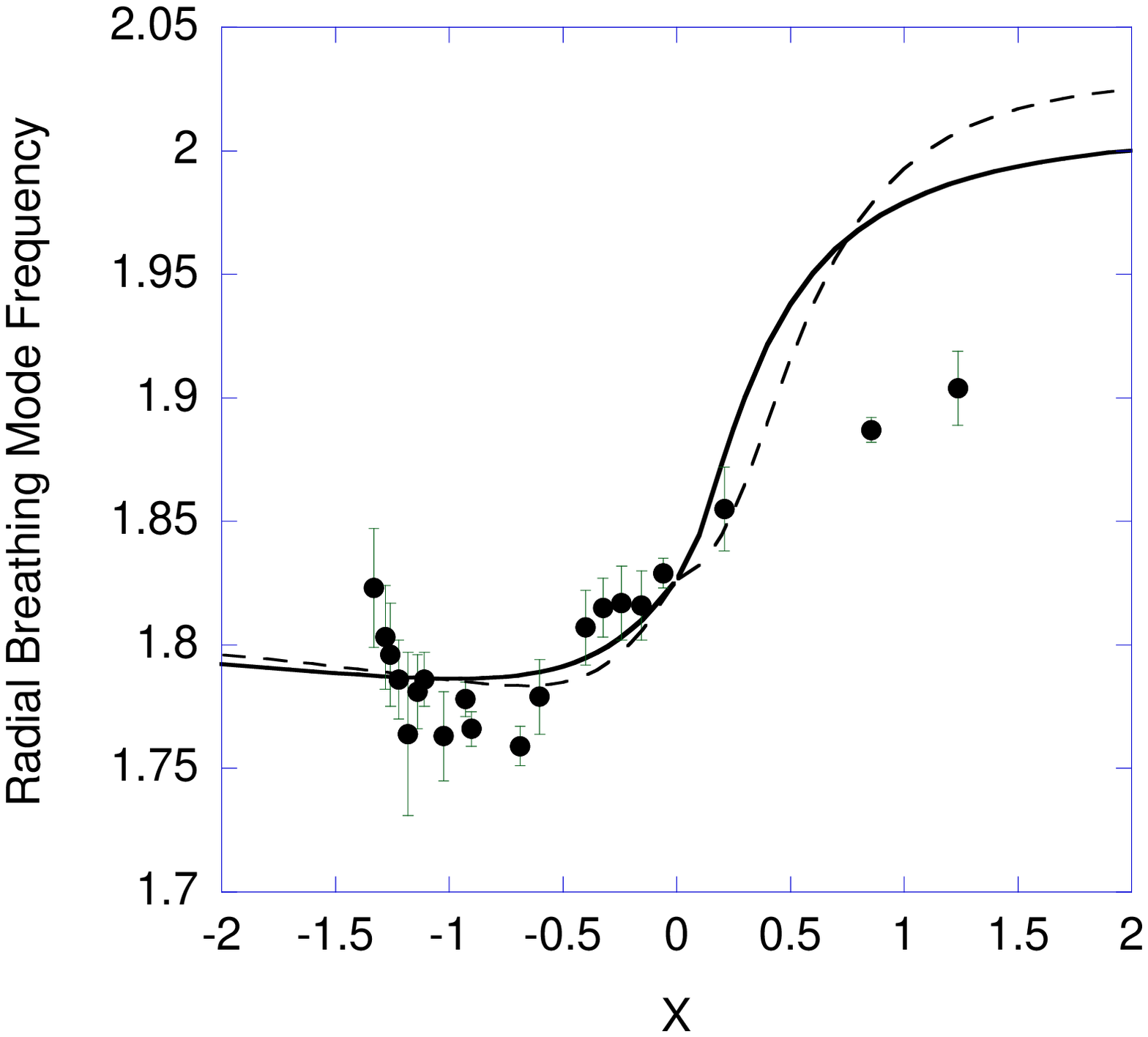}
\end{figure}
Fig. 1.
 Radial breathing mode frequency $\omega_+$  in units of $\omega_{\perp}$ as a
function
 of the dimensional parameter $X=(N^{1/6} a/a_{ho})^{-1}$. In the unitary limit,
$a\rightarrow -\infty $, one expect $\omega/\omega_{\perp}=\sqrt{10/3}
\approx 1.826$. The  solid line  and the dashed line  represent the
 hydrodynamical
 upper bounds calculated using for the energy per particle the [2/2] Pad\'{e}
approximation
 of Refs.[20,21]
and the parameterization of Ref.[28], respectively. The  circular dots
 with error
 bars are the experimental results given by the Duke University group [17].

\pagebreak
\begin{figure}[ht]
\includegraphics{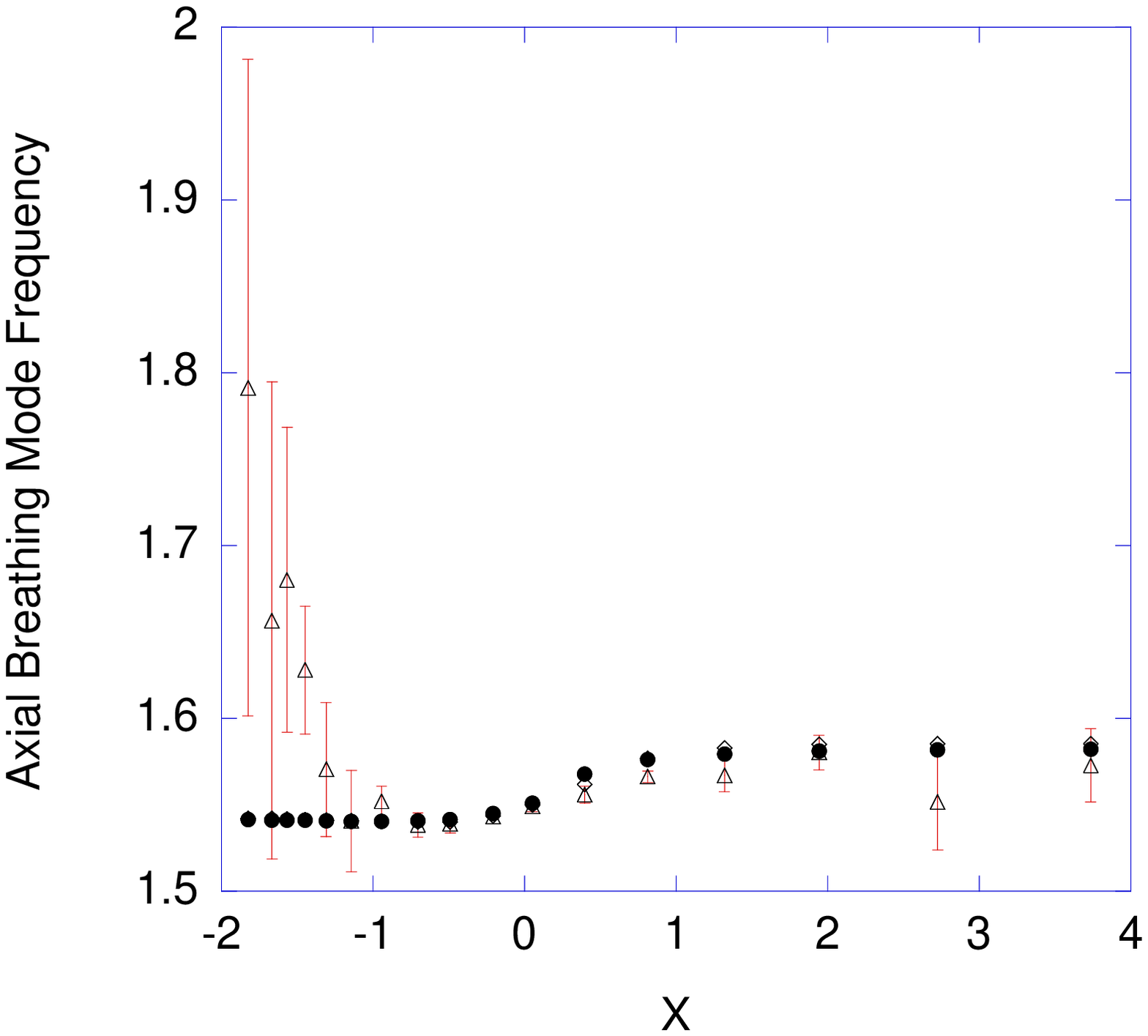}
\end{figure}
Fig.2.
Axial breathing mode  frequency $\omega_-$  in units of $\omega_z$ as a function
of the dimensional parameter $X=(N^{1/6} a/a_{ho})^{-1}$.
The circular dots and diamonds
represent the hydrodynamical
 upper bounds calculated using for the energy per particle the [2/2] Pad\'{e}
approximation
 of Refs.[20,21] and the parameterization of Ref.[28], respectively.
The  triangles   with error
 bars are the experimental results given by the Innsbruck group [16].

\pagebreak
{\bf References}
\vspace{8pt}

\noindent
1. M.E. Gehm  and J. E. Thomas,
Am. Sci. {\bf 92}, 238 (2004).

\noindent
2. C.A. Regal, M. Greiner, and D.S. Jin, Phys. Rev. Lett. {\bf 92}, 040403 
(2004).

\noindent
3. M. Greiner, C.A. Regal, and D.S. Jin, Nature {\bf 426}, 537 (2003).

\noindent
4. K.E. Stecker, G.B. Patridge, and R.G. Hulet, Phys. Rev. Lett. {\bf 91},
 080406 (2003).

\noindent
5. J. Cubizolles, T. Bourdel, S.J.J.M.F. Kokkelmans, G.V Shlyapnikov, and
 C. Salomon, Phys. Rev. Lett. {\bf 91}, 240401 (2003).

\noindent
6. S. Jochim, M. Bartenstein, A. Altmeyer, G. Hendl, C. Chin, J.H. Denschlag,
and
R. Grimm, Phys. Rev. Lett. {\bf 91}, 240402 (2003).

\noindent
7. S. Jochim, M. Bartenstein, A. Altmeyer, G. Hendl, S. Riedl, C. Chin,
J.H. Denschlag, and  R. Grimm, Science {\bf 302}, 2101 (2003).

\noindent
8. M.W. Zwierlein, C.A. Stan, C.H. Schunck, S.M.F. Raupach, S. Gupta,
Z. Hadzibabic, W. Ketterle, Phys. Rev. Lett. {\bf 91}, 250401 (2003).

\noindent
9. C.A. Regal and D.S. Jin, Phys. Rev. Lett. {\bf90}, 230404 (2003).

\noindent
10. K.M. O'Hara, S.L. Hemmer, M.E. Gehm, S.R. Granade, and  J.E. Thomas,
Science {\bf 298}, 2179 (2002).

\noindent
11. J. Kinast, A. Turlapov, J. E. Thomas, Q. Chen, J. Stajic, and
 K. Levin, Science {\bf 307}, 1296 (2005).

\noindent
12. J. Kinast, A. Turlapov, J. E. Thomas, cond-mat/0502507.

\noindent
13. M. Bartenstein, A. Altmeyer, S. Riedl, S. Jochim, C. Chin, J.
Hecker Denschlag, and R. Grimm, Phys. Rev. Lett. {\bf92}, 120401 (2004).

\noindent
14. J. Kinast, S. L. Hemmer, M. E. Gehm, A. Turlapov, and J. E. Thomas,
Phys. Rev. Lett. {\bf 92}, 150402 (2004).

\noindent
15. T. Bourdel, L. Khaykovich, J. Cubizolles, J. Zhang, F. Chevy,
 M. Teichmann, L. Tarruell, S. J. J. M. F. Kokkelmans, and C. Salomon,
Phys. Rev. Lett. {\bf 93}, 050401 (2004)

\noindent
16. M. Bartenstein, A. Altmeyer, S. Riedl, S. Jochim, C. Chin,
 J. Hecker Denschlag, and R. Grimm,  Phys. Rev. Lett. {\bf 92}, 203201 (2004);
cond-mat/0412712.

\noindent
17. J. Kinast, A. Turlapov, and J.E. Thomas, Phys. Rev. A {\bf 70}, 051401(R)
 (2004).

\noindent
18. C. Chin, M. Bartenstein, A. Altmeyer, S. Riedl, S. Jochim,
J. Hecker Denschlag, and  R. Grimm, Science {\bf 305}, 1128 (2004).

\noindent
19.
M. W. Zwierlein, C. A. Stan, C. H. Schunck, S. M. F. Raupach, A. J. Kerman,
 and W. Ketterle, Phys. Rev. Lett. {\bf 92}, 120403 (2004).

\noindent
20. Y.E. Kim and A.L. Zubarev, Phys. Lett. A{\bf 327}, 397 (2004).

\noindent
21. Y.E. Kim and A.L. Zubarev,
 Phys. Rev. A
 {\bf 70}, 033612 (2004).

\noindent
22. Y.E. Kim and A.L. Zubarev, cond-mat/0408283.

\noindent
23. S. Stringari, Europhys. Lett. {\bf 65}, 749 (2004).

\noindent
24. H. Heiselberg, cond-mat/0403041.

\noindent
25. C. Menotti, P.Pedri and S. Stringari, Phys. Rev. Lett. {\bf 89}, 250402
(2002).

\noindent
26. A. Bulgak and G.F. Bertsch, Phys. Rev. Lett. {\bf 94}, 070401 (2005).

\noindent
27. Hui Hu, A. Minguzzi, Xia-Ji Liu, and M. P. Tosi,
Phys. Rev. Lett. {\bf 93}, 190403 (2004)

\noindent
28.  N. Manini and L. Salasnich, con-mat/0407039.

\noindent
29. M. Cozzini and S. Stringari, Phys. Rev. Lett. {\bf 91},
 070401 (2003).

\noindent
30. Y.E. Kim and A.L. Zubarev, Phys. Rev. A {\bf 67}, 015602 (2003).

\noindent
31. Y.E. Kim and A.L. Zubarev, Phys. Rev. A {\bf 69}, 023602 (2004).

\noindent
32.  W. Lenz, Z. Phys. {\bf 56}, 778 (1929); K. Huang and C.N. Yang,
Phys. Rev. {\bf105}, 767 (1957);
T.D. Lee and C.N. Yang, ibid {\bf 105}, 1119 (1957);
V.N. Efimov and M.Ya. Amus'ya, Zh. Eksp. Teor. Fiz. {\bf 47}, 581 (1964) 
[Sov. Phys. JETP {\bf 20},388 (1965)].

\noindent
33.  G.A. Baker, Jr., Int. J. Mod. Phys. B{\bf15}, 1314 (2001); Phys. Rev. C{\bf
60},
054311 (1999).

\noindent
34. J. Carlson, S.-Y. Chang, V.R. Pandharipande, and K.E. Schmidt,
Phys. Rev. Lett. {\bf
 91},
050401 (2003).

\noindent
35. R. Combescot, Phys. Rev. Lett. {\bf 91}, 120401 (2003).

\noindent
36. G. E. Astrakharchik, J. Boronat, J. Casulleras, and and S. Giorgini,
Phys. Rev. Lett. {\bf 93}, 200404 (2004).

\noindent
37. T.L. Ho and E.J. Mueller, Phys. Rev. Lett. {\bf 92}, 160404 (2004);
T.L. Ho, {\it ibid}. {\bf 92}, 090402 (2004).

\noindent
38. G.M. Bruun, Phys. Rev. A {\bf 70}, 053602 (2004).

\noindent
39.  A.J. Leggett, in {\it Modern Trends in the Theory of Condensed
Matter},
 edited by A. Pekalski and R. Przystawa, Lecture Notes in Physics
Vol. 115 (Springer-Verlag, Berlin, 1980) pp. 13-27;
P. Nozi\`{e}res and S. Schmitt-Rink, J. Low. Temp. Phys. {\bf 59}, 195 (1985).

\noindent
40. D. S. Petrov, C. Salomon, and G. V. Shlyapnikov,
Phys. Rev. Lett.{\bf 93}, 090404 (2004).

\noindent
41. R. Combescot and X. Leyronas, cond-mat/0407388.

\noindent
42. E.A. Hylleraas and B. Undheim, Z. Phys. {\bf 65}, 759 (1930).

\noindent
43. For the BEC the first term of the expansion (15) was considered in A.L.
Fetter, J. Low. Temp. Phys. {\bf 106}, 643 (1997).

\noindent
44.  R. Combescot and X. Leyronas, Phys. Rev. Lett. {\bf 93},
138901 (2004).

\noindent
45. L.P. Gorkov and T.K. Melik-Bakhudarov, Sov. Phys. JETP, {\bf 13}, 1018
(1961).

\noindent
46. S. Y. Chang, V. R. Pandharipande,
J. Carlson, and
K. E. Schmidt,
Phys. Rev. A {\bf70}, 043602 (2004).

\noindent
47. M. Bartenstein, A. Altmeyer, S. Riedl, R. Geursen, S. Jochim, C. Chin,
 J. Hecker Denschlag, R. Grimm, A. Simoni, E. Tiesinga, C. J. Williams, and
 P. S. Julienne cond-mat/0408673.

\noindent
48. The measurements of the effective pairing gap [18] support a very small
 value of the ratio (16) for the axial mode case at $B=910$G.
\end{document}